\documentclass[twoside,11pt]{article}

\usepackage[cp1251]{inputenc}
\usepackage{graphicx}

\usepackage{amssymb}
\usepackage{amsmath}
\usepackage{amsfonts}
\usepackage{wasysym}

\begin{document}           

\title{\Large 
Production of matter in the universe via after-GUT interaction}
\author{V.E. Kuzmichev, V.V. Kuzmichev\\[0.5cm]
\itshape Bogolyubov Institute for Theoretical Physics,\\
\itshape National Academy of Sciences of Ukraine, Kiev, 03680 Ukraine}

\date{}

\maketitle

\begin{abstract}

In this paper we propose a model of production of ordinary and dark matter in the decay of a hypothetical 
antigravitating medium in the form of a condensate of (zero-momentum) spinless massive particles 
(denoted as $\phi$)
which fills the universe. The decays of $\phi$-particles into baryons, leptons, and dark matter particles 
are caused by some (after-GUT) interaction with the mass scale between the electroweak and grand 
unification. The observed dark energy is identified with a portion of a condensate which has not decayed 
up to the instant of measurement.
The decay rate of $\phi$-particles $\Gamma_{\phi}$ is expressed through the three parameters - the 
coupling constant  $\alpha_{X}$, the mass scale $M_{X}$ which defines the mass of $X$-particle as the 
mediator of after-GUT interaction, and the energy imparted to the decay products.
We show that the masses of dark matter particle $m_{\chi}\approx 5$ GeV and $\phi$-particle 
$m_{\phi}\approx 15$ GeV can be extracted from the 7-year WMAP and other astrophysical data 
about the contributions of baryon, dark matter, and dark energy  densities to the total matter-energy 
density budget in our universe. 
Such a mass of light WIMP  dark matter agrees with the recent observations of CoGeNT, DAMA, and CDMS. 
The obtained masses of $\phi$- and dark matter particle are concordant with the 
coupling constant of after-GUT interaction $\alpha_{X} \sim \frac{1}{70}$ at 
$M_{X} \sim 6 \times 10^{10}$ GeV, and the decay rate 
$\Gamma_{\phi} \approx 2 \times 10^{-18}\,\mbox{s}^{-1}$. 
The cross-sections of the reactions in which dark matter particles can be produced are calculated.
\end{abstract}

PACS numbers: 98.80.Qc, 98.80.Cq, 95.35.+d, 95.36.+x 

\ \\

\textbf{1. Introduction}
\ \\ [0.3cm]

The present data of modern cosmology pose the principle question about the origin and nature of 
mass-energy constituents of our universe.
The 7-year WMAP and other data \cite{WMAP7} indicate that: the total energy density parameter 
$\Omega_{tot} = 1.006 \pm 0.006$, i.e. the spatial geometry of the observed part of our universe is close
to flat, but it might have a slight positive curvature; the expansion of the universe is accelerating due to 
the action of dominating substance called dark energy; the universe is now comprised of one third dark 
matter, which provides the formation of observed structures com\-posed of baryons and leptons, 
and two thirds dark energy. It is 
remarkable that observed mass of stars is negligibly small (it accounts for $\apprle 0.5\%$ from the total 
amount of mass-energy in the universe \cite{Col,PS}).  In view of the current dominance of dark energy 
over all other forms of matter it is reasonable to consider a model in which ordinary and dark matter 
be decay products of a portion of an antigravitating medium, which fills the universe, 
under the action of some interaction
with gauge coupling between the electroweak and grand unification scales. This interaction leads to the 
violation of CP-invariance which can cause the baryon asymmetry of the universe.
This process is characterized by a very small decay rate and the undecayed part of antigravitating 
medium can be considered as dark energy.
The antigravitating medium itself can be identified with a scalar field called quintessence or 
can be related to the inflaton field which is used in inflation models \cite{AT}.

Previously, we have proposed the quantum cosmological model \cite{K5} in which a hypothetical 
antigravitating medium is a 
condensate of quantized primordial scalar field. We have started with the Einstein-Hilbert action of 
general relativity with matter. The formalism is applied for the case of the homogeneous, isotropic and 
closed universe. The scalar field serves as a surrogate for the matter content of the universe. We pass 
from classical to quantum description using the Dirac's canonical quantization method \cite{Dir}, in which 
the constraints become operators that annihilate physical state vectors, and employing the notion of the so-called 
material reference system (related to relativistic elastic media introduced by DeWitt \cite{DW,BM}). After 
quantization the scalar field changes into a condensate whose properties are described by the quantum 
theory. The direct calculations show that, for quantum states with the mass of a condensate which 
exceeds significantly the Planck mass, a condensate acquires the properties of an antigravitating 
medium with the vacuum-type equation of state. If one discards the quantum effects, a condensate turns 
into an aggregate of separate macroscopic bodies with zero pressure which is known as dust and used 
to model ordinary matter in general relativity. 
The existence of this limit argues, first of all, in favour of reliability of proposed quantum model \cite{K5} 
and, secondly, exhibits the quantum nature of antigravitating property of dark energy.

In this paper we consider the nonstationary universe in which a condensate of a primordial scalar field is 
a source of matter in the form of baryons, leptons, and dark matter, while observed dark energy is a 
portion of a condensate which has not decayed up to the instant of observation. In Section~2 the decay 
rate $\Gamma_{\phi}$ of  particle of a condensate ($\phi$-particle) is derived under assumption of the 
existence of a new force mediated by new virtual massive $X$-particles with the coupling at the mass 
scale $M_{X}$ between the electroweak $M_{W}$ and grand unification $M_{G}$. 
In Section~3 the numerical values of unknown parameters, such as the mean decay rate  
$\overline{\Gamma}_{\phi}$, masses of $\phi$-particle and dark matter particle, are calculated. In 
Section~4 these values are used to obtain the numerical estimation for the  mass scale $M_{X}$ and 
corresponding coupling constant $\alpha_{X}$.
The estimations for the density of $\phi$-particles surrounded by virtual $X$-particle 
cloud and for the cross-section of the reactions in which dark matter particles can be produced are
given. In Section~5 the conclusion remarks are drawn.

\ \\ 

\textbf{2. Matter production}
\ \\ [0.1cm]

According to modern point of view (see, e.g., \cite{KT}), 
the very early universe (close to the Planck era) was filled with
the relativistic matter with the kinetic energy of motion of its constituent particles (temperature $T$)
which exceeds significantly their rest mass energy, $T \gg m_{max}$, where $m_{max}$ is 
the greatest mass of all the particles. Particles may acquire the mass after the
spontaneous symmetry breaking under the cooling of the expanding universe. When the
temperature falls to the value  $T \ll m_{min}$, where $m_{min}$ is the smallest mass
of the particles, matter in the universe is found to be composed of two components,
radiation and massive particles. In the early universe the relativistic matter itself may
arise as a result of transition of the vacuum energy of some primordial scalar field into
the energy of small quantum oscillations near the equilibrium state corresponding to true or false vacuum.

When applying this model to the study of the evolution of matter in the universe, it is
convenient to represent the relativistic matter in the form of the sum of two components,
one of which is composed of massless particles, while another consists of massive particles.
Massless particles form the radiation with which the reference frame can be associated \cite{K5}.
In the era $T \ll m_{min}$, massive particles produce ordinary and dark matter observed in the universe.

In our approach \cite{K5}, massive component is a condensate of scalar field quanta ($\phi$-particles). 
Their mass $m_{\phi}$ can be expressed through the curvature of the potential energy density of a
primordial scalar field. Stationary states of a condensate are characterized by a mass (energy),
$M_{k} = m_{\phi} (k + \frac{1}{2})$, where $M_{k}$ is the eigenvalue of the operator of mass-energy of
a scalar field in a comoving volume,  $k$ is the number of $\phi$-particles which are bosons. 
A condensate is a chargeless medium (for all types of charges). 
Therefore, $\phi$-particle coincides with its antiparticle, 
$\phi = \bar{\phi}$. The momentum of $\phi$-particle vanishes and its spin is $s_{\phi} = 0$.

Let us suppose that baryons ($n,\,p$), leptons ($e^{-},\,\nu_{e}$), dark matter particle ($\chi$) and their antiparticles ($\bar{n},\,\bar{p},\,e^{+},\,\bar{\nu}_{e},\,\bar{\chi}$) 
are produced via the decays of $\phi$-particles in the processes
\begin{eqnarray}\label{1}
    \phi \rightarrow \chi + \nu + n  \quad \mbox{and then} 
    \quad n \rightarrow  p + e^{-} + \bar{\nu},
\end{eqnarray}
or 
\begin{eqnarray}\label{2}
  \phi = \bar{\phi} \rightarrow \bar{\chi} + \bar{\nu} + \bar{n}   \quad \mbox{and then} 
   \quad \bar{n} \rightarrow \bar{p} + e^{+} + \nu.
\end{eqnarray}

In every separate decay (\ref{1}) or (\ref{2}) the baryon invariance is violated. If both decays (\ref{1}) and 
(\ref{2}) are equiprobable, then the system composed from matter, antimatter (including dark sectors), 
and a fraction of $\phi$-particles of a condensate which have not decayed up to the instant under 
consideration, remains chargeless. The processes (\ref{1}) and (\ref{2}) 
imply that the dark matter particle and antiparticle are scalar (spin $s_{\chi} = 0$) or vector 
($s_{\chi} = 1$) particles. It follows from Ref. \cite{FHZ} that a dark matter particle may be a scalar 
particle with scalar interaction. The presence of the neutrino in the decay (\ref{1}) guarantees the spin 
conservation law. 

The data of astrophysical observations indicate that matter 
dominates over antimatter in the observed part of the universe, pointing to B-violation. 
According to the CPT-theorem all interactions are invariant under the succession of the three operations 
C, P, and T taken in any order. The baryon asymmetry of the universe can be caused by CP-violation. 
Supposing that this principle of quantum field theory also takes place for 
some unknown interaction (we shall call it after-GUT interaction), via which the processes (\ref{1}) and 
(\ref{2}) occur, one can conclude that CP or T-invariance breaks down in the decays (\ref{1}) and (\ref{2}). The 
T-violation is confirmed by the fact that the probability of inverse fusion reaction of three and more 
particles into one zero-momentum particle is negligibly small\footnote{This 
property is well-known from quantum 
scattering theory and chemical reactions where it is shown that triple (and more multiple) simultaneous
collisions are many orders of magnitude less probable than double collisions.}. 
The CP non-invariance of the after-GUT interaction supplies the arrow of time.

Particles and antiparticles in the decay products of the processes (\ref{1}) and (\ref{2}) can annihilate 
between themselves and contribute to the cosmic background radiation.
Net amounts of protons, leptons, and dark matter particles constitute matter in the luminous and dark
forms in the universe.

Taking into account above-mentioned issues, 
let us examine the decay (\ref{1}) in which dark and baryonic matter are produced.
We consider this process using an analogy with neutron and proton decays.
In the standard model, the decay of neutron is the result of weak interaction, mediated by virtual $W$-boson exchange,
\begin{equation}\label{3}
    n \rightarrow  p + W^{-}  \quad \mbox{and then} \quad W^{-} \rightarrow e^{-} + \bar{\nu}.
\end{equation}
The rate $\Gamma_{n}$ of decay of a neutron into the final $pe\nu$-state is equal to
\begin{equation}\label{4}
   \Gamma_{n} = \frac{1}{\tau_{n}} = \alpha_{W}^{2}\,\frac{\Delta m^{5}}{M_{W}^{4}}
\end{equation}
(in units $\hbar = c = 1$), where $\tau_{n}$ is the mean life of neutron, $\Delta m = m_{n}-m_{p}$ is the energy imparted by the $W$-boson to the leptons $e\nu$, $M_{W}$ is the mass of the $W$-boson, $m_{n}$ and $m_{p}$ are the masses of neutron and proton, respectively. The quantity $\alpha_{W}^{2}$ is dimensionless constant which characterises the strength of weak interaction and includes small radiative and other quantum corrections.
This number is close to the value $G_{F} m_{p}^{2} = 1.027 \times 10^{-5}$, where $G_{F} = 1.166 \times 10^{-5}$ GeV$^{-2}$ is the Fermi coupling constant,  
\begin{equation}\label{6}
    \alpha_{W}^{2} = G_{F} m_{p}^{2}\, (1 - \delta) \sim 10^{-5},
\end{equation}
where $\delta \sim O(10^{-1})$ is small correction.

In any grand unified theory (GUT) (see, e.g., the review \cite{SM}) the proton lifetime $\tau_{p}$ is given by the following relation
\begin{equation}\label{8}
     \Gamma_{p} = \frac{1}{\tau_{p}} = \alpha_{G}^{2}\,\frac{m_{p}^{5}}{M_{G}^{4}},
\end{equation}
where $ \Gamma_{p}$ is the decay rate of proton, $\alpha_{G}$ and $M_{G}$ are two 
parameters. Universal gauge coupling  $\alpha_{G}$ is defined at the grand unification scale $M_{G}$. 
The SUSY GUTs give the value $\alpha_{G} \sim \frac{1}{25}$ at $M_{G} \sim 3 \times 10^{16}$ GeV.
These parameters correspond to the proton lifetime equal to $\tau_{p} \sim 10^{37}$ yrs.
We can use the estimation 
\begin{equation}\label{9}
    \alpha_{G}^{2} \sim 10^{-3}.
\end{equation}
It can be compared with $\alpha_{W}^{2}$ from Eq. (\ref{6})
\begin{equation}\label{10}
    \alpha_{W}^{2} \sim 10^{-5} <  \alpha_{G}^{2} \sim 10^{-3}.
\end{equation}
This relation is useful for the estimation of the coupling constant of after-GUT interaction.

Let us suppose that the decay of $\phi$-particle in the process (\ref{1}) is mediated by exchange of virtual $X$-particles which are quanta of some new field with the mass scale
\begin{equation}\label{12}
    M_{X} \gg m_{\chi} > m_{n} \gg m_{\nu},
\end{equation}
where $m_{\chi}$ and $m_{\nu}$ are the masses of dark matter particle and neutrino. 
Generally speaking, a few versions of such a decay are possible 
\begin{eqnarray}\label{13}
    \phi \rightarrow  \chi + X \quad \mbox{and then} \quad X \rightarrow n + \nu ,\nonumber \\
    \phi \rightarrow n + X \quad \mbox{and then} \quad X \rightarrow \chi + \nu ,\nonumber \\
    \phi \rightarrow \nu + X \quad \mbox{and then} \quad X \rightarrow \chi + n .
\end{eqnarray}
In Fig.~1, the diagram represents the first process in Eq. (\ref{13}). The two other diagrams can be obtained by cyclic permutation of the particles $\chi$, $n$, $\nu$.
\begin{figure}[ht]
\begin{center}
\includegraphics*[width=0.4\textwidth]{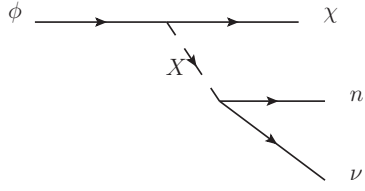}
\end{center}
\caption{The diagram which corresponds to the first decay in (\ref{13}).} \label{fig:1}
\end{figure} 

Since the spin $s_{\phi} = 0$, one concludes that
in the first process in Eq. (\ref{13}) the spins of $\chi$- and $X$-particles can vanish, $s_{\chi} = 0$ and
$s_{X} = 0$, or be equal to unity, $s_{\chi} = 1$ and $s_{X} = 1$. In this case $\chi$- and $X$-particles
are scalars or vector bosons. In the second and third processes in Eq. (\ref{13}) $\chi$-particle
is still a scalar or vector boson, but $X$-particle is a fermion with the spin $s_{X} = \frac{1}{2}$.

We consider the processes (\ref{13}) by analogy to those
described by Eqs. (\ref{4}) and (\ref{8}). We neglect the contribution from integration with respect to intermediate momentum of $X$-particle in corresponding transition amplitudes, as well as contributions into the decay rate of $\phi$-particle $\Gamma_{\phi}$ arising from all higher-order diagrams.
Then $\Gamma_{\phi}$ can be written as follows
\begin{equation}\label{14}
    \Gamma_{\phi} = \alpha_{X}^{2}\, \frac{Q^{5}}{M_{X}^{4}},
\end{equation}
where 
\begin{equation}\label{15}
    Q = m_{\phi} - (m_{\chi} + m_{n} + m_{\nu})
\end{equation}
is the energy imparted by $\phi$-particle at rest to the decay products in (\ref{1}). The dimensionless coupling constant $\alpha_{X}^{2}$ takes into account the contributions into $\Gamma_{\phi}$ from 
the first or both the second and third pole diagrams for the processes (\ref{13}). 
In the case, when $X$-boson has a supersymmetric partner with the same mass ($X$-fermion),
all three pole diagrams contribute into the decay rate $\Gamma_{\phi}$.   

Using Eqs. (\ref{4}), (\ref{8}), and (\ref{14}) one can write the expressions for the ratios of the decay rate of $\phi$-particle to the neutron and proton decay rates
\begin{equation}\label{16}
    \frac{\Gamma_{\phi}}{\Gamma_{n}} =\frac{ \alpha_{X}^{2}}{ \alpha_{W}^{2}}\,
     \left(\frac{Q}{\Delta m}\right)^{5}\,\left(\frac{M_{W}}{M_{X}}\right)^{4}
\end{equation}
and
\begin{equation}\label{17}
   \frac{\Gamma_{\phi}}{\Gamma_{p}} =\frac{ \alpha_{X}^{2}}{ \alpha_{G}^{2}}\,
     \left(\frac{Q}{m_{p}}\right)^{5}\,\left(\frac{M_{G}}{M_{X}}\right)^{4}.
\end{equation} 
Here the quantities $\Gamma_{\phi},\ \alpha_{X}^{2}, \ Q$ and $M_{X}$ are 
unknown. The decay rate $\Gamma_{\phi}$ and the parameter $Q$ can be calculated independently in 
the kinetic theory which considers two-step processes (\ref{1}) and (\ref{2}) as dynamical ones and uses  
the astrophysical data about the contributions of baryon, dark matter, and dark energy densities to the 
total matter-energy density budget in our universe. 

\ \\

\textbf{3. Dark energy and light WIMP dark matter}
\ \\ [0.1cm]

The total energy density of matter in the universe relative to critical density $\Omega_{tot}$ taken at some fixed
instant of the proper time $t$ can be represented by the sum of the terms
\begin{equation}\label{19}
    \Omega_{tot} = \Omega_{B} + \Omega_{L} + \Omega_{DM} + \Omega_{CMB} + \Omega_{DE},
\end{equation}
where $\Omega_{B}$, $\Omega_{L}$, $\Omega_{DM}$, and $\Omega_{CMB}$ are the energy densities 
of baryons, leptons, dark matter particles, and the cosmic microwave background radiation, 
$\Omega_{DE}$ is the density of dark energy (i.e. a portion of a condensate which has not 
decayed up to the instant under consideration). These constituents can be written as follows
\begin{eqnarray}\label{36}
    \Omega_{B} & = & \frac{2m_{p}\delta N_{p}(t)}{a^{3}(t) H^{2}(t)},\quad
    \Omega_{L} = \frac{2m_{l}\delta N_{l}(t)}{a^{3}(t) H^{2}(t)},\nonumber \\ 
    \Omega_{DM} & = & \frac{2m_{\chi}\delta N_{\chi}(t)}{a^{3}(t) H^{2}(t)},\quad
    \Omega_{DE} = \frac{2m_{\phi}\delta N_{\phi}(t)}{a^{3}(t) H^{2}(t)},
\end{eqnarray}
where 
$a$ is the cosmic scale factor taken in unit of the Planck length $l_{P} = \sqrt{\frac{2G}{3\pi}}$, 
$m_{l}$ is the sum of masses of all leptons in the final state of reaction (\ref{1}),  and $H$ is the 
dimensionless Hubble expansion rate (in unit of time $t_{P} = l_{P}$)\footnote{In
units under consideration, the value $\frac{1}{2} a^{3}$ is the dimensionless comoving volume of the universe, while
$H^{2}$ corresponds to the dimensionless critical density.}. 
All masses in the equations (\ref{36})
are taken in units of the Planck mass  $m_{P} = l_{P}^{-1}$.
The quantities
\begin{equation}\label{371}
    \delta N_{i} = N_{i} - N_{\bar{i}}, \qquad i = \{p,l,\chi\},
\end{equation}
are the differences between the number of particles $N_{i}$ and antiparticles $N_{\bar{i}}$ of $i$th type
as the functions of $t$,
$m_{i}\delta N_{i}$ is the total mass of particles of $i$th type that remained after annihilation,
$\delta N_{\phi}$ is the number of undecayed $\phi$-particles which form dark energy.

According to the quantum model description \cite{K5}, when the universe expands, the following 
condition is realized at every instant of time for large enough number of the $\phi$-particles,
\begin{equation}\label{20}
    \langle a \rangle_{k} = M_{k}, 
\end{equation}  
where $\langle a \rangle_{k}$ is the mean value of the scale factor $a$ in the k-state of the universe with 
the mass of a condensate $M_{k}$. The equation (\ref{20}) can be interpreted as a mathematical 
formulation of the Mach's principle proposed by Sciama \cite{Di,K8}. 
In the classical limit, the evolution of the mean value $\langle a \rangle_{k}$ in time is described by the 
Einstein-Friedmann equations. The observed part of 
the present-day universe has the `radius' $a \sim 10^{28}$ cm,
and its total mass can be estimated as $M \sim 10^{80}$ GeV. In dimensionless units it corresponds
to the condition $a \sim M \sim 10^{61}$, which agrees with Eq. (\ref{20}).

Under the expansion in accordance with the Hubble law,
$(d/dt)a(t) = H(t) a(t)$,
the mass of a condensate in the universe changes as follows\footnote{In the standard cosmological 
model with the cosmological constant, the similar relation takes place. During the expansion the vacuum 
energy density remains constant $\rho_{\Lambda} = \mbox{const}$, but the energy $M_{\Lambda} = \frac{1}{2} a^{3} \rho_{\Lambda}$ evolves as $\frac{d M_{\Lambda}}{dt} = 3 H M_{\Lambda}$.}
\begin{equation}\label{23}
    \frac{dM_{k}(t)}{dt} = H(t) M_{k}(t).
\end{equation}
Since $\phi$-particles are indistinguishable and their masses do not change with time, 
the same equation holds for the number of undecayed particles, 
\begin{equation}\label{231}
     \frac{d\delta N_{\phi}(t)}{dt} = H(t) \delta N_{\phi}(t).
\end{equation}

We suppose that $\phi$-particle, neutron, and proton decay independently with some decay rates 
$\Gamma_{\phi} = \Gamma_{\bar{\phi}}$, $\Gamma_{n} = \Gamma_{\bar{n}}$, 
and $\Gamma_{p} = \Gamma_{\bar{p}}$. 
The $\phi$-particles are supposed to be unstable ones. Then, taking into account two-channel 
$\phi$-particle decay as in the schemes (\ref{1}) and (\ref{2}), 
we can write the equations of the evolution of amounts of unstable particles in the form
\begin{eqnarray}\label{24} 
    \frac{d\delta N_{\phi}(t)}{dt} = -\,\lambda (t)\delta N_{\phi}(t), \quad
     \frac{d\delta N_{n}(t)}{dt} = -\,\Gamma _{n}(t)\delta N_{n}(t) + 
                                 \Gamma _{\phi}(t)\delta N_{\phi}(t), \nonumber \\
     \frac{d\delta N_{p}(t)}{dt} = -\,\Gamma _{p}(t)\delta N_{p}(t) +
        \Gamma _{n}(t)\delta N_{n}(t),
\end{eqnarray}
where
\begin{equation}\label{25}
    \lambda(t) = \Gamma_{\phi}(t) - H(t)
\end{equation}
is an effective decay rate which takes into account the decay of $\phi$-particles and the increase in
the mass of a condensate during the expansion of the universe. According to the equations
(\ref{24}), the number of neutrons (protons) decreases as a result of their decay and increases due to
the decay of $\phi$-particles (neutrons).
Since $\Gamma _{p} \ll \Gamma _{n}$, one can take $\Gamma _{p} = \Gamma _{\bar{p}} = 0$. Choosing the initial conditions as follows
\begin{equation}\label{26}
    \delta N_{\phi}(t') = N, \quad \delta N_{n}(t') = 0, \quad \delta N_{p}(t') = 0,
\end{equation}
where $N$ is the number of $\phi$-particles at initial instant of time $t'$ prior to emergence of
protons and neutrons in the system\footnote{For our universe $t' \sim 10^{-6}$ s.}, 
we find the solution of the set (\ref{24})
\begin{equation}\label{27}
    \frac{\delta N_{\phi}(t)}{N} = e^{- \overline{\lambda}\Delta t},
\end{equation}
\begin{equation}\label{28}
    \frac{\delta N_{n}(t)}{N} = \int_{t'}^{t}\!\!dt_{1} \Gamma_{\phi}(t_{1})\,
e^{- \int_{t'}^{t_{1}}\!\!dt_{2} \lambda(t_{2})}\,
e^{- \int_{t_{1}}^{t}\!\!dt_{2} \Gamma_{n}(t_{2})},
\end{equation}
\begin{equation}\label{29}
    \frac{ \delta N_{p}(t)}{N} = \int_{t'}^{t}\!\!dt_{1} \Gamma_{n}(t_{1})
\int_{t'}^{t_{1}}\!\!dt_{2} \Gamma_{\phi}(t_{2})\,
e^{- \int_{t'}^{t_{2}}\!\!dt_{3} \lambda(t_{3})}\,
e^{- \int_{t_{2}}^{t_{1}}\!\!dt_{3} \Gamma_{n}(t_{3})},
\end{equation} 
where $\overline{\lambda} = \overline{\Gamma}_{\phi} - \overline{H}$ and
\begin{equation}\label{30}
    \overline{\Gamma}_{\phi} = \frac{1}{\Delta t}
\int_{t'}^{t}\!\!dt_{1} \Gamma_{\phi}(t_{1}), \qquad
     \overline{H} = \frac{1}{\Delta t} \int_{t'}^{t}\!\!dt_{1} H(t_{1})
\end{equation}
are the mean decay rate of $\phi$-particles and the mean Hubble expansion rate on the time interval 
$\Delta t = t - t'$.

The decay rate of $\phi$-particles $\Gamma_{\phi}(t)$ is unknown. We assume that the decay rates 
$\Gamma_{\phi}$ and $\Gamma_{n}$ depend very weakly on averaging interval $\Delta t$. 
Then from Eqs. (\ref{28}) 
and (\ref{29}) we obtain
\begin{equation}\label{31}
     \frac{\delta N_{n}(t)}{N} = 
\frac{\overline{\Gamma}_{\phi}}{\overline{\Gamma}_{n} - \overline{\lambda}}\,
\left(e^{-\overline{\lambda}\Delta t} - e^{-\overline{\Gamma}_{n}\Delta t}\right),
\end{equation} 
\begin{equation}\label{32}
     \frac{ \delta N_{p}(t)}{N} = \frac{\overline{\Gamma}_{\phi}}{\overline{\lambda}}\,
\left[1 + \frac{1}{\overline{\Gamma}_{n} - \overline{\lambda}}\, 
\left(\overline{\lambda}\,e^{-\overline{\Gamma}_{n}\Delta t} -  
\overline{\Gamma}_{n}\,e^{-\overline{\lambda}\Delta t}  \right) \right].
\end{equation}

The equations (\ref{27}), (\ref{31}) and (\ref{32}) are consistent with the conservation law of particles,
\begin{equation}\label{321}
   \delta N_{\phi}(t) + \delta N_{n}(t) + \delta N_{p}(t) = N + \frac{\overline{H}}{\overline{\lambda}}\, N \left(1 -  e^{- \overline{\lambda}\Delta t}\right),
\end{equation}
where the second summand on the right-hand side takes into account the number of $\phi$-particles which have appeared in the universe under its expansion and then decayed.

In our universe today $\overline{\Gamma}_{n} = 1.12 \times 10^{-3}$ s$^{-1}$, 
$H_{0} = 71.0 \pm 2.5 \,\mbox{km}\, \mbox{s}^{-1}\mbox{Mpc}^{-1}$ 
and the age $t_{0} = 13.75 \pm 0.13$ Gyr \cite{WMAP7}. 
Taking $\Delta t = t_{0}$ for estimation we find that
\begin{equation}\label{33}
    H_{0}\Delta t = 0.999, \qquad \overline{\Gamma}_{n}\Delta t = 4.86 \times 10^{14}. 
\end{equation}
We have supposed that the decay of $\phi$-particle is caused by the action of the after-GUT interaction 
with the mass scale $M_{X}$. Then the inequality $\overline{\Gamma}_{\phi} \ll \overline{\Gamma}_{n}$ 
must hold, and
\begin{equation}\label{34}
    \overline{\lambda} \ll \overline{\Gamma}_{n}.
\end{equation}
Under this condition, the number of baryons $\delta N_{p}$ in the expanding universe obeys the law
\begin{equation}\label{35}
    \frac{\delta N_{p}(t)}{N} = \frac{\overline{\Gamma}_{\phi}}{\overline{\lambda}}\,
     \left[1 - e^{-\overline{\lambda}\Delta t} \right].   
\end{equation}
The ratio $\frac{\Omega_{B}}{\Omega_{DE}}$ can be written as follows
\begin{equation}\label{38}
    \frac{\Omega_{B}}{\Omega_{DE}} = \sqrt{\frac{g_{p}}{g_{\phi}}}\,\,
\frac{\overline{\Gamma}_{\phi}}{\overline{\lambda}}\,
 \left[e^{\overline{\lambda}\Delta t } - 1\right],
\end{equation}
where $g_{p} = G m_{p}^{2} = 0.59 \times 10^{-38}$ and $g_{\phi} = G m_{\phi}^{2}$ are the 
dimensionless gravitational coupling constants for proton and $\phi$-particle, respectively, 
$G = 6.707\times 10^{-39}$  GeV$^{-2}$ is the Newtonian gravitational constant. 

If $\overline{\Gamma}_{\phi} \gg \overline{H}$, then $\overline{\lambda} \approx \overline{\Gamma}_{\phi}$ and
\begin{equation}\label{39}
     \frac{\Omega_{B}}{\Omega_{DE}} = \sqrt{\frac{g_{p}}{g_{\phi}}}\,\,
 \left[e^{\overline{\Gamma}_{\phi}\Delta t}  - 1\right].
\end{equation}
This limit was considered in Ref. \cite{K1}. 

If $\overline{\Gamma}_{\phi} \ll \overline{H}$, then $\overline{\lambda} \approx - \overline{H}$ and
\begin{equation}\label{40}
    \frac{\Omega_{B}}{\Omega_{DE}} = 0.
\end{equation}
It means that in this case baryons (ordinary matter) are not produced in the universe.

If $\overline{\Gamma}_{\phi} \approx \overline{H}$, and $\overline{\lambda}\Delta t \ll 1$, then
\begin{equation}\label{41}
    \frac{\Omega_{B}}{\Omega_{DE}} =
\sqrt{\frac{g_{p}}{g_{\phi}}}\,\,\overline{\Gamma}_{\phi}\Delta t \left[1 + \frac{1}{2}\, \overline{\lambda}\Delta t + \dots \right].
\end{equation}
The ratio (\ref{41}) describes the case, when the number of $\phi$-particles remains almost unchanged
during the expansion of the universe,
$\delta N_{\phi}(t) \approx N = \mbox{const}$. It means that the amount of a condensate in the universe 
is semipermanent, but its energy density diminishes during the expansion. Then the coupling constant 
$g_{\phi}$ is equal to
\begin{equation}\label{42}
    g_{\phi} = g_{p} \,\left(\frac{\Omega_{DE}}{\Omega_{B} }\,
\overline{H}\Delta t\right)^{2}_{\overline{\lambda}=0}.
\end{equation}

Using the WMAP and other data \cite{WMAP7}, $\Omega_{B} = 0.0456 \pm 0.0016$, 
$\Omega_{DE} = 0.728^{+0.015}_{-0.016}$, and Eq. (\ref{33}), and setting 
$\overline{H} \Delta t \approx H_{0} \Delta t$,  we obtain
\begin{equation}\label{43}
    g_{\phi} \approx 254\, g_{p}.
\end{equation}
Under this assumption, the decay rate of $\phi$-particle is very close to the Hubble expansion rate 
$H_{0}$ and it equals to
\begin{equation}\label{44}
    \overline{\Gamma}_{\phi} \approx 2.3 \times 10 ^{-18}\,\mbox{s}^{-1}.
\end{equation} 
This value is close to the value found in Ref. \cite{K1}, where Eq. (\ref{231}) was not taken into account. 
That case is equivalent to the limit $\overline{\Gamma}_{\phi} \gg \overline{H}$. Therefore the values 
(\ref{43}) and (\ref{44}) may be considered as realistic, since they have changed very slightly and seem 
almost model-independent.

The coupling constant (\ref{43}) corresponds to the mass
\begin{equation}\label{45}
    m_{\phi} \approx 16\, m_{p}.
\end{equation} 

Taking into account that, according to the decay schema (\ref{1}), 
the number of dark matter $\chi$-particles  is almost equal to the number of baryons,
\begin{equation}\label{48}
    \delta N_{\chi}(t)\approx \delta N_{p}(t),
\end{equation}
from Eq. (\ref{36}) we find that
\begin{equation}\label{49}
    \frac{\Omega_{B}}{m_{p}} \approx \frac{\Omega_{DM}}{m_{\chi}}.
\end{equation}

For the observed values of $\Omega_{B}$ and $\Omega_{DM} = 0.227 \pm 0.014$ \cite{WMAP7} it follows that
\begin{equation}\label{50}
    \frac{\Omega_{DM}}{\Omega_{B}} \approx 5.
\end{equation}
With the regard for the errors of measurement of $\Omega_{DM}$ and $\Omega_{B}$, one can take the value of the mass of $\chi$-particle equal to
\begin{equation}\label{51}
    m_{\chi} \approx 5\, m_{p}.
\end{equation}
This value is in the range $m_{\chi} \sim 1 - 10$ GeV indicated in Ref. \cite{FHZ}. 
It agrees with the observations of CoGeNT \cite{CoG}, DAMA \cite{DAMA}, and CDMS \cite{CDMS}.
The equation (\ref{48}) is a relationship between the dark matter and baryon chemical potentials with the precise value $c_{1} = 1$ of the coefficient $c_{1}$ introduced in ADM models (cf. Ref. \cite{FHZ}).

The values of masses (\ref{45}) and (\ref{51}) show that the decay (\ref{1}) occurs with the release of energy $Q\approx 10\,m_{p}$ in the form of kinetic energy of decay products.

From the equations (\ref{36}) it follows
\begin{equation}\label{54}
    \frac{\Omega_{DE}}{\Omega_{DM}} = \frac{m_{\phi} \delta N_{\phi}}
        {m_{\chi}\delta N_{\chi}}.
\end{equation}
Using the observed values of $\Omega_{DE}$ and $\Omega_{DM}$, and the obtained values of the masses $m_{\phi}$ (\ref{45}) and $m_{\chi}$ (\ref{51}), we find that
\begin{equation}\label{55}
    \delta N_{\phi} \approx \delta N_{\chi}.
\end{equation}
It means that the chemical potentials of dark energy and dark matter coincide. The equation (\ref{55}) is a manifestation of the so-called coincidence problem, $\frac{\Omega_{DE}}{\Omega_{DM}} \approx 3.2$. 

In conclusion to this Section, we note that the low mass dark matter problem can be analyzed, for instance, within the context of the standard model with scalar dark matter, ADM models, or the minimal supersymmetric standard model with neutralino dark matter (see, e.g., the bibliographies in Refs. \cite{FHZ,CoG,AA,FZN,BDFS}).
A different model based on the quantum cosmological approach involving the available data on the abundances of baryons and dark matter in our universe was proposed in Ref. \cite{K1}. Obtained restriction on the mass of dark matter particle $m_{\chi} < 15$ GeV, with the preference for $m_{\chi} \sim 5 - 10$ GeV, agrees with the values given in Refs. \cite{FHZ,CoG} and in this paper. 

\ \\

\textbf{4. Estimations of coupling constant and mass scale}
\ \\ [0.1cm]

Let us suppose that the decay rate of $\phi$-particle (\ref{14}) is of the same order of magnitude as the 
mean decay rate (\ref{44}). For the present epoch we have 
\begin{equation}\label{56}
    \frac{\Gamma_{\phi}}{\Gamma_{p}} \sim 10^{27}, \qquad
    \frac{\Gamma_{\phi}}{\Gamma_{n}} \sim 2 \times 10^{-15}.
\end{equation}
It is interesting that the ratio
\begin{equation}\label{57}
    \frac{\Gamma_{n}}{\Gamma_{p}} \sim 5 \times 10^{41}
\end{equation}
gives the value, which is of the same order of magnitude as the well-known Eddington's magic numbers. 
It can be used for a more precise definition of the proton lifetime.

From Eqs. (\ref{14}) and (\ref{44}), for $Q \approx 10 \, m_{p}$, we obtain
\begin{equation}\label{58}
    M_{X} \approx \sqrt{\alpha_{X}}\,0.5 \times 10^{12}\,m_{p}.
\end{equation}
The same formula follows from Eqs. (\ref{16}) and (\ref{17}). 
The equation (\ref{58}) shows a relation between the mass scale $M_{X}$ and the coupling constant 
$\alpha_{X}$.

The coupling constant $ \alpha_{X}$ must satisfy the inequality
\begin{equation}\label{59}
      \alpha_{W}^{2} \sim 10^{-5} <  \alpha_{X}^{2} < \alpha_{G}^{2} \sim 10^{-3}.
\end{equation}
It gives the value for the mass scale $M_{X}$ which lies in the interval
\begin{equation}\label{60}
    0.3 < (M_{X} \times 10^{-11}\,m_{p}^{-1} ) < 0.9,
\end{equation}
and we can accept the following numerical values
\begin{equation}\label{61}
   \alpha_{X} \sim \frac{1}{70} \quad \mbox{at} \quad 
    M_{X} \sim 6 \times 10^{10} \, \mbox{GeV} 
\end{equation}
for the parameters of after-GUT interaction. The radius of action of the after-GUT force is
\begin{equation}\label{62}
    R_{X} = M_{X}^{-1} \sim 3 \times 10^{-25}\,\mbox{cm}.
\end{equation}
It gives  the following value 
\begin{equation}\label{63}
    n_{\phi} \sim M_{X}^{3} \sim 10^{73} \,\mbox{cm}^{-3}
\end{equation}
for the density of the dark energy quasiparticles considered as $\phi$-particles 
surroun\-ded by virtual $X$-particle cloud. The $\phi$-particle is massive and it can, in principle, exhibit 
itself through the gravitational action, but its gravitational coupling constant 
$g_{\phi} \approx 1.5 \times 10^{-36}$ is very small. The after-GUT coupling 
$\alpha_{X} \sim 10^{-2}$ has the same order of magnitude as the fine structure constant 
$\alpha \approx \frac{1}{137}$. This allows to consider the subtle processes 
which occur via virtual $X$-particle exchange as more realistic for identification of dark matter and dark energy.

Let us consider the reaction
\begin{equation}\label{631}
\bar{\nu} + \phi \rightarrow  \chi + n,
\end{equation}
in which dark matter particles are produced.
In Fig.~2 the diagram of this reaction is shown.
\begin{figure}[ht]
\begin{center}
\includegraphics*[width=0.4\textwidth]{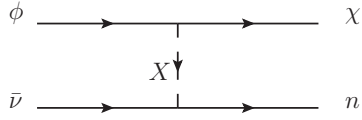}
\end{center}
\caption{The diagram which describes the reaction (\ref{631}).} \label{fig:2}
\end{figure} 

The corresponding cross-section is equal to  
\begin{equation}\label{64}
    \sigma(\bar{\nu}\phi) \simeq \frac{\Gamma_{\phi}}{v M_{X}^{3}},
\end{equation}
where $v$ is the relative $\bar{\nu}-\phi$ motion velocity and the denominator is the incident flow of 
antineutrinos in the rest frame of $\phi$-particle. Since in this reference frame the velocity $v$ is equal to 
the velocity of the incident antineutrino, we can take $v \approx c$, and for the decay rate (\ref{44}) and 
the mass scale $M_{X}$ from (\ref{61}) we obtain
\begin{equation}\label{65}
     \sigma(\bar{\nu}\phi) \sim 0.7 \times 10^{-74}\,\mbox{GeV}^{-2}.
\end{equation}

Dark matter particles can be produced in another reaction as well,
\begin{equation}\label{651}
\bar{n} + \phi \rightarrow  \chi + \nu.
\end{equation}
For nonrelativistic antineutrons the cross-section of this process is
\begin{equation}\label{652}
    \sigma(\bar{n}\phi) \simeq \frac{0.7}{\sqrt{E_{\bar{n}}}} \frac{\Gamma_{\phi}}{M_{X}^{3}},
\end{equation}
where $E_{\bar{n}}$ is the kinetic energy of antineutron taken in GeV. For thermal and fast antineutrons
the cross-sections are equal to
\begin{eqnarray}\label{653}
     \sigma(\bar{n}\phi) & \sim & 10^{-69}\,\mbox{GeV}^{-2} \quad \mbox{for} \quad
     E_{\bar{n}} = 0.25 \times 10^{-10}\, \mbox{GeV}, \nonumber \\
      \sigma(\bar{n}\phi) & \sim & 1.6 \times 10^{-73}\,\mbox{GeV}^{-2} \quad \mbox{for} \quad
     E_{\bar{n}} = 10^{-3}\, \mbox{GeV},
\end{eqnarray}
respectively. 
These values are much less than an elastic scattering cross-section of dark matter particle with
nucleus $\sigma \sim 10^{-13}$ GeV$^{-2}$ used for the estimation of a mass of dark matter particle
\cite{FHZ,CoG}.

\ \\

\textbf{5. Conclusion remarks}
\ \\ [0.1cm]

Thus, one can conclude that the observed values of the densities $\Omega_{B}$, $\Omega_{DM}$, and 
$\Omega_{DE}$ in the model of the decays (\ref{1}) and (\ref{2}) lead to the values of mass of dark 
energy particle $\sim 15$ GeV and mass of dark matter particle $\sim 5$ GeV. 
The obtained masses of dark energy and dark matter particles are consistent with the parameters of 
after-GUT interaction (\ref{61}) and the decay rate $\Gamma_{\phi}$ (\ref{44}). The parameters of after-
GUT interaction are in so-called \textit{gauge desert} - the domain between the electroweak and grand 
unification scales.

The density of the dark energy particles (\ref{63}) shows that, after space-averaging, the volume 
$\sim (2 \,\mbox{m})^{3}$ contains the same number of dark energy particles as the number of 
equivalent baryons $\sim 10^{80}$ in the observed part of our universe.
The cross-sections of the reactions (\ref{631}) and (\ref{651}) via virtual $X$-particle 
exchange are very small, but they are finite.

Within the framework of quark model of baryons, one may conclude that a condensate (and, hence, dark 
energy) can be a chargeless aggregate of point-like quarks and gluons, a kind of quark-gluon plasma. 
Then $\phi$-particles will be the particle-like excitations of this plasma. The existence of 
virtual $X$-particles gives the possibility for $\phi$-particles to decay with 
a very small probability into the observed luminous and dark matter. 
As regards $X$-particle, it can be one of the supersymmetric particles \cite{SM}. 

Using the hypothesis about the existence of a new scale $M_{X} \sim 10^{10}$ GeV,
a more general scheme of unification of now four (without gravitation) 
fundamental interactions into one single force with a scale $M_{X} \sim 10^{16}$ GeV may be proposed 
in the framework of quark model of baryons. In the simplest version of a new 
theory some new gauge group $G'$ must be included formally into the scheme
\begin{equation}\label{66}
    G \stackrel{G}{\longrightarrow} G' \stackrel{X}{\longrightarrow} SU(3) \times SU(2) \times U(1) \stackrel{W}{\longrightarrow} SU(3) \times U(1),
\end{equation}
where $G$ is some group which realizes the local symmetry principle and contains the group
$G'$ as a subgroup. The model with such a group $G$ must have three mass scales - the
masses of gauge particles $G$, $X$, and $W$. The values of these masses $M_{G}$, $M_{X}$, 
and $M_{W}$ characterize the spontaneous symmetry breaking of $G$ to $G'$, 
then to $SU(3) \times SU(2) \times U(1)$ and finally to $SU(3) \times U(1)$.
Since  $M_{G} \gg M_{X} \gg M_{W}$, then there is a vast hierarchy of gauge symmetries.
In order to build a grand unification theory which will realize the scheme
(\ref{66}), it is necessary to make the group $G$ specific.

\end{document}